\documentclass[%
 reprint,
 amsmath,amssymb,
 aps,
prb,
floatfix,
]{revtex4-2}

\usepackage{graphicx}
\usepackage{dcolumn}
\usepackage{subfigure}
\usepackage{bm}
\usepackage{hyperref}
\usepackage[T2A,T1]{fontenc}
\usepackage[utf8]{inputenc}
\usepackage{float}

\usepackage{amsmath,amssymb}
\usepackage{lmodern}
\usepackage{mathtools}
\usepackage{bm}
\usepackage{physics}
\usepackage{xcolor}
\usepackage{graphicx}
\usepackage{calc}

\usepackage{hyperref}
\hypersetup{
    colorlinks=true,
    linkcolor=red,
    citecolor=red,
    filecolor=magenta,      
    urlcolor=blue,
    }

\begin{document}

\title{Resonant state expansion for acoustic resonators. \\ Part II. Scattering problem}

\author{Egor~Domoratskii$^{1,2}$}

\author{Vladimir~Igoshin$^{2}$}

\author{Nikolay Solodovchenko$^{1,2,3}$}

\author{Mingzhao~Song$^{1}$}

\author{Yong~Li$^{2,4}$}
\email{yongli@tongji.edu.cn}

\author{Mihail~Petrov$^{2}$}
\email{m.petrov@metalab.ifmo.ru}

\author{Andrey Bogdanov$^{1,3}$}
\email{a.bogdanov@hrbeu.edu.cn}

\affiliation{$^{1}$ Qingdao Innovation and Development Center, Harbin Engineering University, Qingdao 266000, China}

\affiliation{$^{2}$ School of Physics and Engineering, ITMO University, St. Petersburg 197101, Russia}

\affiliation{$^{3}$ Ioffe Institute, St. Petersburg 194021, Russia}

\affiliation{$^{4}$ Institute of Acoustics, School of Physics Science and Engineering, Tongji University, Shanghai 200092, China}

\begin{abstract}
We develop a resonant-state expansion formulation for acoustic scattering by individual resonators. The scattered pressure and particle-velocity fields are expanded over the resonant states of the system, with excitation amplitudes determined by overlap integrals between the incident field and the resonant states over the resonator volume. Using the acoustic energy flux, we derive expressions for the extinction, scattering, and absorption cross-sections and show that the extinction spectrum can be resolved into contributions from individual resonant states. The formulation is first validated for a homogeneous two-dimensional cylinder, where it reproduces the analytical Mie-theory solution. We then consider a sectorally perturbed cylinder with coupled azimuthal modes and demonstrate agreement with finite-element simulations. Finally, we combine the eigenvalue and scattering formulations for a material-programmed hard-wall annular metaatom and reproduce its scattering spectra and near fields. The developed framework provides a physically transparent modal approach to acoustic scattering by open resonators with reduced symmetry and spatially structured material parameters.

\end{abstract}

\maketitle

\section{Introduction}

Acoustic scattering by resonators, inhomogeneities, and structured media is central to the control of sound in phononic crystals and acoustic metamaterials~\cite{lu2009phononic}. Resonant scattering underlies sound absorption~\cite{huang2023sound}, compact acoustic resonators supporting bound states in the continuum~\cite{deriy2022bound}, and non-Hermitian acoustic systems with exceptional points~\cite{igoshin2024exceptional}. Similar scattering phenomena play an important role in labyrinthine and space-coiling structures, where subwavelength channels and engineered material distributions provide strong control over acoustic-wave propagation~\cite{liang2013_space-coiling,maurya2016double,yin2026design,timankova2026experimental}.

For canonical geometries, acoustic scattering can be treated accurately using separation of variables and partial-wave expansions. The corresponding solutions are naturally expressed in cylindrical or spherical harmonics and provide exact or semi-analytical results for circular, spherical, concentric, and related geometries~\cite{williams1999fourier}. A more general framework is provided by the T-matrix method, which relates the expansion coefficients of an incident field to those of the scattered field. Its foundations are closely connected with multiple-scattering theory~\cite{waterman1961multiple}, and the method was formulated for acoustic scattering in the seminal work of Waterman~\cite{waterman1969new}. The relation between T-matrix formulations and boundary-integral methods has also been studied in detail~\cite{martin2003connections}, and a broader overview of acoustic T-matrix methods was given by Waterman~\cite{waterman2009tmatrix}. Modern implementations provide flexible numerical tools for acoustic scattering calculations, including the recent \texttt{acoustotreams} package~\cite{ustimenko2026acoustotreams}.

Despite its broad applicability, the T-matrix depends on frequency and must generally be evaluated separately at each frequency of a spectral sweep. Moreover, expansions in spherical or cylindrical waves can require large truncation orders and may become numerically challenging for scatterers with strongly elongated or flattened shapes, complex internal structures, or strongly reduced symmetry. These limitations become particularly relevant for modern acoustic metaatoms containing internal channels, angular sectors, inclusions, or spatially varying material parameters. Full-wave numerical methods remain applicable in such cases, but repeated frequency sweeps and parameter studies may require substantial computational effort and do not directly reveal how individual resonances contribute to the measured scattering spectrum.

A complementary description of open acoustic scattering is based on the resonant states (RSs) of the corresponding open system~\cite{domoratskii2026resonant-I,laude2023quasinormal}. Unlike the normal modes of a closed Hermitian resonator, RSs satisfy outgoing-wave boundary conditions and possess complex eigenfrequencies that encode both the resonance frequency and radiative decay~\cite{laude2023quasinormal,vial2024quasinormal}. Modal approaches therefore provide a natural connection between the internal resonant dynamics of an open system and its response to external excitation~\cite{alpeggiani2017quasinormal,lobanov2018resonant}. Acoustic coupled-mode and modal formulations have been used to reconstruct scattering matrices, transmission spectra, and forced responses of open resonators and cavities~\cite{maksimov2015coupled,tong2017modal,tong2017forced}. Related quasinormal-mode approaches have been developed for radiating phononic and elastic systems~\cite{laude2023quasinormal,vial2024quasinormal}.

The relation between open-system modes and observable scattering characteristics has been developed particularly systematically in electromagnetism. The scattering matrix can be expanded directly in terms of quasinormal modes~\cite{alpeggiani2017quasinormal}. Within the resonant-state expansion (RSE), the Green's function of an open system is represented using its RSs together with any additional continuum or nonresonant contributions required by its analytic structure~\cite{muljarov2011brillouin,doost2013resonant,doost2014resonant}. Among the subsequent developments of the RSE, particularly relevant here are its extension to dispersive and generalized electromagnetic media~\cite{muljarov2016resonant,muljarov2018resonant} and its application to the direct calculation of scattering matrices, electromagnetic fields, and scattering cross-sections~\cite{lobanov2018resonant,lobanov2019resonant}.

An important advantage of the resonant-state description is the possibility of resolving the driven fields into contributions from individual resonances. In acoustics, both the scattered pressure and particle-velocity fields can be expanded linearly over the RSs. This modal decomposition also carries over directly to the extinction, which is linear in the scattered field, so that the extinction spectrum can be resolved into contributions from individual RSs. By contrast, scattering and absorption contain quadratic field products and therefore generally include interference terms between different resonances. The RSE thus provides information that is not accessible from the total spectra alone, allowing particular spectral features to be associated with specific modes, their symmetry, and their mutual interference. Despite the successful application of RSE-based scattering formulations in electromagnetism, an analogous systematic treatment remains considerably less developed in acoustics.

In the present work (Part~II), we develop the scattering formulation of the acoustic RSE based on the general framework established in Part~I of this series~\cite{domoratskii2026resonant-I}. The scattered pressure and velocity fields are expanded over the normalized RSs, while their excitation amplitudes are expressed through overlap integrals involving the incident field and the material perturbation. From these fields, we derive the extinction, scattering, and absorption cross-sections. Since the extinction is linear in the scattered field, its spectrum can be decomposed into contributions from individual RSs, making it possible to identify the modes responsible for particular spectral features and to analyze their symmetry, spectral positions, radiative properties, and interference.

The rest of the paper is organized as follows. In Sec.~II, we formulate the acoustic scattering problem within the RSE and derive the expansion of the scattered pressure and velocity fields together with the corresponding modal excitation coefficients. In Sec.~III, we establish the energy-balance relations and derive the extinction, scattering, and absorption cross-sections, including their decomposition into contributions from individual RSs. In Sec.~IV, we validate the approach for a homogeneous two-dimensional cylinder against the analytical Mie-theory solution and then consider a sectorally perturbed cylinder, where different azimuthal orders are coupled and separation of variables is no longer applicable. In Sec.~V, we combine the eigenvalue and scattering formulations for a material-programmed hard-wall annular metaatom, where the principal geometry is incorporated into the reference RS basis, and the cell-wise material distribution is introduced through the RSE.


\section{Resonant-State Expansion for Acoustic Scattering Problem}
\label{sec:waves_scattering}

In general, the linear acoustic equations for complex pressure $p(\mathbf{x})$ and velocity $\mathbf{v}(\mathbf{x})$ fields in media  have the following form:
\begin{equation}
    \begin{cases}
        i \omega \beta_\omega(\mathbf{x}) p(\mathbf{x}) = \nabla\cdot\mathbf{v}(\mathbf{x}), \\
        i \omega \displaystyle \rho_\omega(\mathbf{x}) \mathbf{v}(\mathbf{x}) = \nabla p(\mathbf{x}) + \mathbf{f}(\mathbf{x}),
    \end{cases}
    \label{eq:acoustic_equation_system}
\end{equation}
where $\beta_\omega(\mathbf{x})$ and $\rho_\omega(\mathbf{x})$ are the material's compressibility and mass density with frequency dispersion, denoted by index $\omega$, respectively; $\mathbf{f}(\mathbf{x})$ is the force volume density~\cite{landau1987fluid, toftul2019acoustic}. The connection between material density and compressibility is given by the sound velocity $c_\omega(\mathbf{x}) = [\rho_\omega(\mathbf{x}) \beta_\omega(\mathbf{x})]^{-1/2}$. We consider time-dependent parts of pressure and velocity fields as $\exp(-i \omega t)$ with a complex frequency $\omega$.

We use matrix representation of Eq.~\eqref{eq:acoustic_equation_system} in the following form
\begin{equation}
    \hat{\mathbb{D}}(\mathbf{\mathbf{x}})\vec{\mathbb{F}}(\mathbf{x}) = \omega \hat{\mathbb{P}}_\omega(\mathbf{x})\vec{\mathbb{F}}(\mathbf{x}) - \vec{\mathbb{J}}(\mathbf{x}),
    \label{eq:acoustic_matrix_equation}
\end{equation}
given by the matrix operator $\hat{\mathbb{D}}(\mathbf{\mathbf{x}})$, the material parameters matrix $\hat{\mathbb{P}}_\omega(\mathbf{x})$, acoustic field vector $\vec{\mathbb{F}}(\mathbf{x})$ and force source vector $\vec{\mathbb{J}}(\mathbf{x})$ as follows
\begin{equation*}
    \begin{aligned}
        \hat{\mathbb{D}}(\mathbf{\mathbf{x}}) =
    \begin{bmatrix}
        0                   & i\nabla_\mathbf{x} \cdot \\
        -i\nabla_\mathbf{x} & 0
    \end{bmatrix},
    &\quad
    \hat{\mathbb{P}}_\omega(\mathbf{x}) =
    \begin{bmatrix}
        -\beta_\omega(\omega, \mathbf{x})   & 0 \\
        0                                   & \rho_\omega(\omega, \mathbf{x})
    \end{bmatrix},
    \\
    \vec{\mathbb{F}}(\mathbf{x}) =
    \begin{bmatrix}
        p(\mathbf{x}) \\
        \mathbf{v}(\mathbf{x})
    \end{bmatrix},
    &\quad
    \vec{\mathbb{J}}(\mathbf{x}) = 
    \begin{bmatrix}
        0 \\
        - i\mathbf{f}(\mathbf{x})
    \end{bmatrix}.
    \end{aligned}
\end{equation*}

The solution of Eq.~\eqref{eq:acoustic_matrix_equation} can be found by introducing dyadic Green's function (GF) integrated over the whole system volume $V$ with the force density source as follows
\begin{equation}
    \vec{\mathbb{F}}_\omega(\mathbf{x}) = \int_V d\mathbf{x}^\prime \hat{\mathbb{G}}_\omega(\mathbf{x}, \mathbf{x}') \vec{\mathbb{J}}(\mathbf{x}^\prime),
    \label{eq:solution_by_green_function}
\end{equation}
The dyadic GF satisfies the equation $\left[ \omega \hat{\mathbb{P}}_\omega(\mathbf{x}) - \hat{\mathbb{D}}(\mathbf{x}) \right] \hat{\mathbb{G}}_\omega(\mathbf{x}, \mathbf{x}') = \hat{\mathbb{I}} \delta(\mathbf{x} - \mathbf{x}')$, coming from Eq.~\eqref{eq:acoustic_matrix_equation} with unit tensor $\hat{\mathbb{I}}$. The GF can be represented as a sum over its resonant poles if it is a meromorphic
\begin{equation}
    \hat{\mathbb{G}}_\omega(\mathbf{x}, \mathbf{x}^\prime)
    =
    \sum_n
    \frac{
        \vec{\mathbb{F}}_n(\mathbf{x})
        \vec{\mathbb{F}}_n^\mathsf{T}(\mathbf{x}^\prime)
    }{
        \omega-\omega_n
    },
    \label{eq:green_function_expansion}
\end{equation}
where $\vec{\mathbb{F}}_n(\mathbf{x})$ and $\omega_n$ are the eigenvectors and eigenvalues, respectively, of the homogeneous problem
\begin{equation}
    \hat{\mathbb{D}}\vec{\mathbb{F}}_n(\mathbf{x})
    =
    \omega_n
    \hat{\mathbb{P}}_{\omega_n}(\mathbf{x})
    \vec{\mathbb{F}}_n(\mathbf{x}).
    \label{eq:matrix_equation_homogeneous}
\end{equation}
However, even for simple open geometries, the GF generally possesses branch-cut singularities in addition to its discrete poles. These contributions must be included in the RSE to obtain a complete spectral representation~\cite{domoratskii2026resonant-I}.

We consider the acoustic system consisting as a background medium with an acoustic resonator as a perturbation. The background media material parameters include density $\rho_{b,\omega}(\mathbf{x})$ and compressibility $\beta_{b,\omega}(\mathbf{x})$. In contrast, the resonator material parameters $\Delta \rho_\omega(\mathbf{x})$ and $\Delta \beta_\omega(\mathbf{x})$ have zero value everywhere except in the resonator region. Thus, it material parameters matrix operator is $\hat{\mathbb{P}}_\omega(\mathbf{x}) = \hat{\mathbb{P}}_{b,\omega}(\mathbf{x}) + \Delta \hat{\mathbb{P}}_\omega(\mathbf{x})$. In the following, we restrict our analysis to non-dispersive media and, therefore, omit the frequency index $\omega$ from the material parameters.

In general, one may find scattering problem solution by introducing the total acoustic field $\vec{\mathbb{F}}(\mathbf{x})$ consisting of background field $\vec{\mathbb{F}}_b(\mathbf{x})$ incident on resonator and scattered field $\vec{\mathbb{F}}_s(\mathbf{x})$ as follows $\vec{\mathbb{F}}(\mathbf{x}) = \vec{\mathbb{F}}_b(\mathbf{x}) + \vec{\mathbb{F}}_s(\mathbf{x})$, where $\vec{\mathbb{F}}_{b,s}(\mathbf{x}) = [p_{b,s}(\mathbf{x}), \mathbf{v}_{b,s}(\mathbf{x})]$ and both components of acoustic field are summarized separately. Substitution of the total field expansion into the background and scattered into Eq.~\eqref{eq:acoustic_matrix_equation} gives us the following inhomogeneous equation for the scattered field $\vec{\mathbb{F}}_s(\mathbf{x})$ (see details in Supplemental Material~\cite{Supplement_II}, Sec.~I):
\begin{equation}
    \hat{\mathbb{D}}(\mathbf{\mathbf{x}}) \vec{\mathbb{F}}_s(\mathbf{x}) =
    \omega \hat{\mathbb{P}}_\omega(\mathbf{x}) \vec{\mathbb{F}}_s(\mathbf{x}) - \vec{\mathbb{J}}(\mathbf{x}),
    \label{eq:scattered_field_equation}
\end{equation}
where $\vec{\mathbb{J}}(\mathbf{x}) = - \omega \Delta \hat{\mathbb{P}}(\mathbf{x}) \vec{\mathbb{F}}_b(\mathbf{x})$.

We obtain the solution to the equation~\eqref{eq:scattered_field_equation} within RSE framework substituting the GF from Eq.~\eqref{eq:solution_by_green_function} and use scattered field expansion
\begin{equation}
    \vec{\mathbb{F}}_s(\mathbf{x},\omega) = \sum_n \alpha_n(\omega) \vec{\mathbb{F}}_n(\mathbf{x}),
    \label{eq:scattered_field}
\end{equation}
and the GF expansion into the RSs $\vec{\mathbb{F}}_n(\mathbf{x})$ given by Eq.~\eqref{eq:green_function_expansion} to show (see details in Supplemental Material~\cite{Supplement_II}, Sec.~I) that excitation coefficients $\alpha_n$ are
\begin{equation}
    \alpha_n(\omega) = - \frac{\omega}{\omega - \omega_n} \int_V d\mathbf{x}^\prime \vec{\mathbb{F}}_n^\mathsf{T}(\mathbf{x}^\prime) \Delta \hat{\mathbb{P}}_\omega(\mathbf{x}^\prime) \vec{\mathbb{F}}_b(\mathbf{x}^\prime).
    \label{eq:excitation_coefficients}
\end{equation}
That relation is referred to as the overlap integral and shows the interaction between the incident wave and acoustic resonator eigemode $\vec{\mathbb{F}}_n(\mathbf{x})$ with corresponding eigenfrequency $\omega_n$ at defined frequency $\omega$.

\begin{figure} [t]
    \centering
    \includegraphics[width =\linewidth]{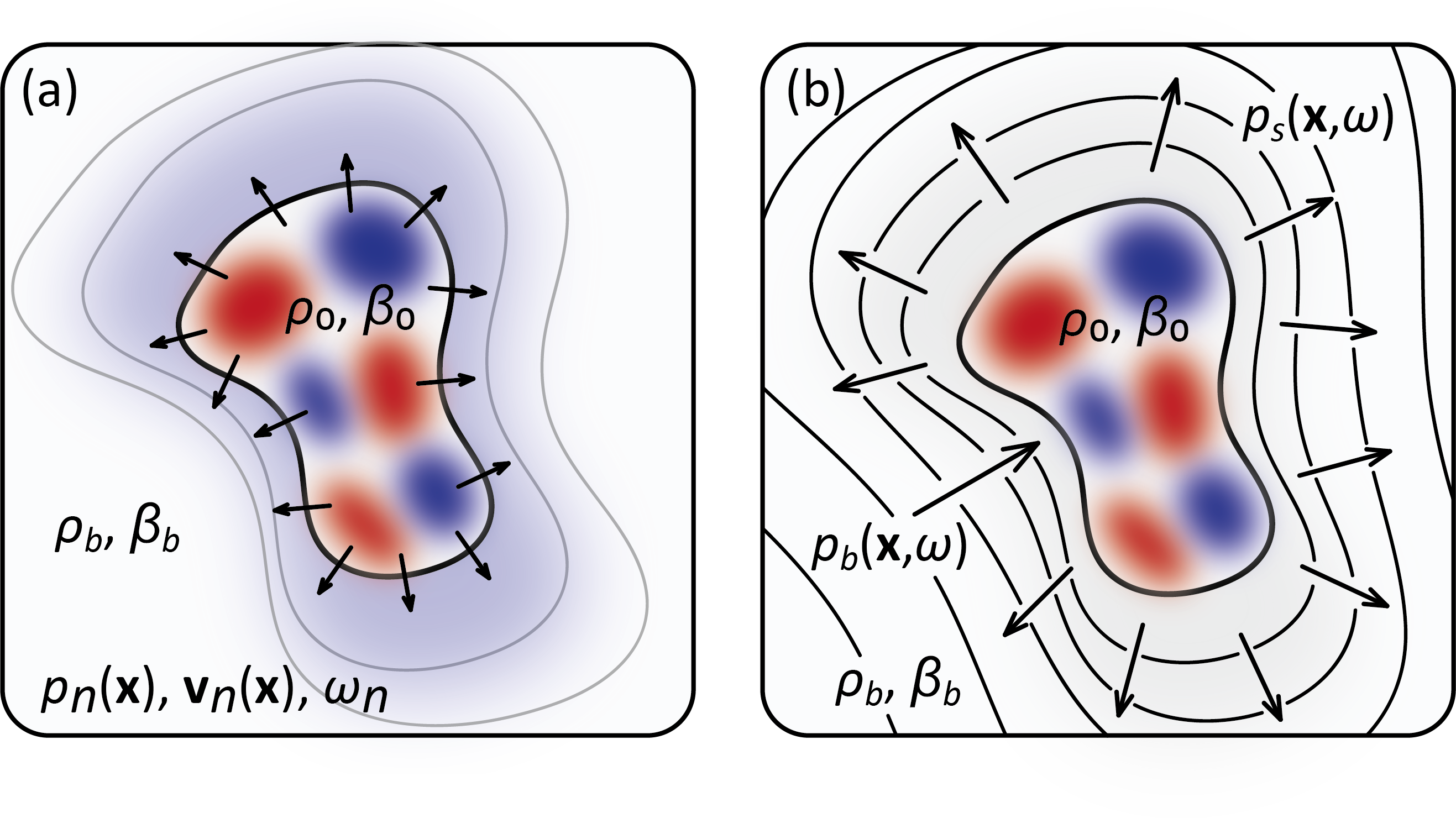}
    \caption{
    (a) Pressure field $p_n(\mathbf{x})$ RSs module distribution (maxima and minima are shown by dark red and dark blue, respectively) for eigenfrequency $\omega_n$ in the considered arbitrary resonator (grey shaded area) with material parameters $\rho_0$ and $\beta_0$ in the background media with parameters $\rho_b$ and $\beta_b$. The resonator has the outgoing boundary condition revealed by outgoing black arrows and a blue shaded area.
    (b) Background pressure field $p_b(\mathbf{x}, \omega)$ incident on the resonator and induces the scattered pressure field $p_s(\mathbf{x}, \omega)$ with direction shown by arrows and dark grey shaded area.
    }
    \label{Draft_Figure_1}
\end{figure}

Thus, the scattering problem solution is found by the following two steps, one after another, as shown in Fig.~\ref{Draft_Figure_1}. In the first step, we use Eq.~\eqref{eq:matrix_equation_homogeneous} to calculate the eigenfrequencies $\omega_n$ and eigenvectors $\{p_n(\mathbf{x}), \mathbf{v}_n(\mathbf{x})\}$ of a resonator with arbitrary shape and material parameters (mass density $\rho_0$ and compressibility $\beta_0$), embedded in a background medium characterized by $\rho_b$ and $\beta_b$. Once eigenfrequencies $\omega_n$ and eigenvectors $\{ p_n(\mathbf{x}), \mathbf{v}_n(\mathbf{x})\}$ of the considered system are calculated analytically or numerically, it is possible to consider background pressure field $p_b(\mathbf{x}, \omega)$ incident on the resonator and find scattered pressure field $p_s(\mathbf{x}, \omega)$ expanded into RSs with overlap integrals, given by Eq.~\eqref{eq:excitation_coefficients}.

\section{Scattering Observables and Modal Contributions}
\label{sec:energy_transfer}

We assume that the background medium is lossless, such that
$\mathrm{Im}(\rho_b) = \mathrm{Im}(\beta_b) = 0$. The time-averaged acoustic energy flux is defined as
\begin{equation}
    \mathbf{\Pi}(\mathbf{x})
    =
    \frac{1}{2}
    \mathrm{Re}\!\left[
        p^*(\mathbf{x})\mathbf{v}(\mathbf{x})
    \right],
\end{equation}
which represents the acoustic analogue of the Poynting vector~\cite{tsimokha2022acoustic, wiley2006poynting, burns2020poynting, bliokh2019poynting}. The power absorbed by the acoustic resonator is then defined as the negative outward flux of the total acoustic intensity through a closed surface $S$ enclosing the resonator,
\begin{equation}
    P_\mathrm{abs}
    =
    -\oint_S
    \mathbf{\Pi}(\mathbf{x})\cdot d\mathbf{S}.
    \label{eq:absorbed_power}
\end{equation}

Introducing the decomposition of the total acoustic field into the incident background and scattered fields,
$p=p_b+p_s$ and $\mathbf{v}=\mathbf{v}_b+\mathbf{v}_s$, the total flux can be separated into the scattered and interference contributions. Since the background medium is lossless, the net flux of the incident field through the closed surface $S$ vanishes. Accordingly, the absorbed, scattered, and extinction powers satisfy the energy-balance relation
\begin{equation}
    P_\mathrm{ext}=P_s+P_\mathrm{abs},
    \label{eq:power_balance}
\end{equation}
where
\begin{equation}
    \begin{aligned}
        P_s
        &=
        \frac{1}{2}
        \oint_S
        \mathrm{Re}
        \left(
            p_s^*\mathbf{v}_s
        \right)
        \cdot d\mathbf{S},
        \\
        P_\mathrm{ext}
        &=
        -\frac{1}{2}
        \oint_S
        \mathrm{Re}
        \left(
            p_b^*\mathbf{v}_s
            +
            p_s^*\mathbf{v}_b
        \right)
        \cdot d\mathbf{S}.
    \end{aligned}
    \label{eq:power_surface_integrals}
\end{equation}

Using the divergence theorem together with the acoustic equations~\eqref{eq:acoustic_equation_system}, the surface integrals in Eq.~\eqref{eq:power_surface_integrals} can be transformed into volume integrals. Writing the material parameters as
$\rho=\rho_b+\Delta\rho$ and $\beta=\beta_b+\Delta\beta$, one obtains
\begin{equation}
    P_\mathrm{abs}
    =
    \frac{\omega}{2}
    \int_V
    d\mathbf{x}\,
    \left[
        \mathrm{Im}(\Delta\rho)
        |\mathbf{v}|^2
        +
        \mathrm{Im}(\Delta\beta)
        |p|^2
    \right],
    \label{eq:absorbed_power_volume_integral}
\end{equation}
where the integrand is nonzero only within the region of the material perturbation, i.e. scatterer. The extinction power can be written as
\begin{equation}
    P_\mathrm{ext}
    =
    \frac{\omega}{2}
    \mathrm{Im}
    \int_V
    d\mathbf{x}\,
    \left[
        \Delta\rho\,
        \mathbf{v}_b^*\cdot\mathbf{v}
        +
        \Delta\beta\,
        p_b^*p
    \right],
    \label{eq:extinction_power_volume_integral}
\end{equation}
whereas the scattered power is
\begin{equation}
    P_s
    =
    -\frac{\omega}{2}
    \mathrm{Im}
    \int_V
    d\mathbf{x}\,
    \left[
        \Delta\rho\,
        \mathbf{v}_s^*\cdot\mathbf{v}
        +
        \Delta\beta\,
        p_s^*p
    \right].
    \label{eq:scattered_power_volume_integral}
\end{equation}
Equations~\eqref{eq:absorbed_power_volume_integral}--\eqref{eq:scattered_power_volume_integral} satisfy the energy-conservation relation~\eqref{eq:power_balance}. In particular, for a lossless scatterer, $\mathrm{Im}(\Delta\rho)=\mathrm{Im}(\Delta\beta)=0$, so that $P_\mathrm{abs}=0$ and $P_\mathrm{ext}=P_s$. In case of lossy background medium, i.e.,
$\mathrm{Im}(\rho_b)\neq0$ and/or $\mathrm{Im}(\beta_b)\neq0$, an additional background-dissipation contribution must be taken into account. The general expressions for a lossy background are given in Supplemental Material~\cite{Supplement_II}.

The powers given by Eqs.~\eqref{eq:absorbed_power_volume_integral}--\eqref{eq:scattered_power_volume_integral} are evaluated numerically using the resonant-state expansion of the scattered field in Eq.~\eqref{eq:scattered_field}, with the excitation coefficients defined by Eq.~\eqref{eq:excitation_coefficients}. Because the extinction power is linear in the scattered field, substitution of the modal expansion into Eq.~\eqref{eq:extinction_power_volume_integral} yields a direct decomposition of $P_\mathrm{ext}$ into contributions associated with individual resonant states. In contrast, the scattered and absorbed powers contain quadratic field products and therefore generally include interference terms between different resonant states.

Once the absorbed, scattered, and extinction powers are determined, the corresponding cross-sections are obtained by normalizing them to the incident-wave intensity,
\begin{equation}
    I_0=\frac{|p_0|^2}{2Z_b},
\end{equation}
where $p_0$ is the incident pressure amplitude and $Z_b=\rho_b c_b$ is the acoustic impedance of the lossless background medium. Accordingly,
\begin{equation}
    \sigma_\mathrm{abs}
    =
    \frac{P_\mathrm{abs}}{I_0},
    \qquad
    \sigma_s
    =
    \frac{P_s}{I_0},
    \qquad
    \sigma_\mathrm{ext}
    =
    \frac{P_\mathrm{ext}}{I_0}.
    \label{eq:cross_sections}
\end{equation}

\begin{figure} [t]
    \centering
    \includegraphics[width = \linewidth]{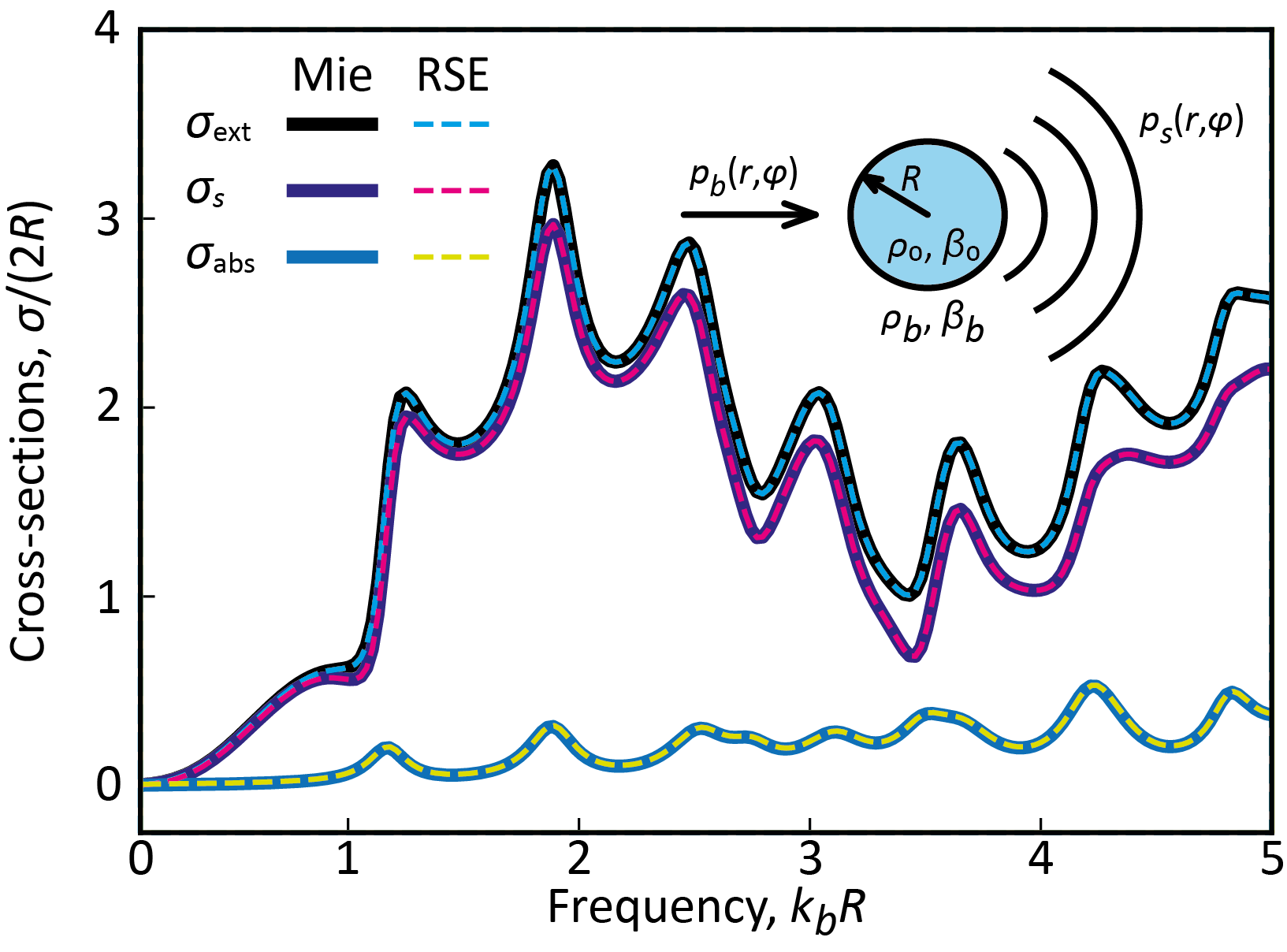}
    \caption{
    The extinction $\sigma_\mathrm{ext}$, absorption $\sigma_\mathrm{abs}$, and scattering $\sigma_s$ cross-section spectra of the homogeneous system obtained within the Mie-theory framework (solid lines) in Supplemental Material~\cite{Supplement_II}, Sec.~III and obtained numerically for all RSs by Eqs.~\eqref{eq:scattered_power_volume_integral} --~\eqref{eq:absorbed_power_volume_integral} (dashed lines). The considered cylindrical structure consists of the open resonator with radius $R$ and material parameters $\rho_0 = 10 \rho_b$, $c_0 = (1  - i \tan\delta_c) c_b/2$ normalized to the background air material parameters $\rho_b = 1.2~\mathrm{kg/m}^3$ and $c_b = 343~\mathrm{m/s}$. Losses are introduced by $\tan\delta_c = 10^{-2}$.
    }
    \label{Draft_Figure_2}
\end{figure}

\section{Applications to 2D Acoustic Resonators}

To begin with, we consider a homogeneous axisymmetric cylinder of radius $R = 10~\mathrm{cm}$ as a two-dimensional analytically solvable reference system. It is surrounded by a homogeneous background medium (air with sound velocity $c_b = 343~\mathrm{m/s}$ and density $\rho_b = 1.2~\mathrm{kg/m}^3$) without any absorption. We introduce losses in the system only for the sound velocity of the resonator by the parameter $\tan\delta_c$ as $c_0 = (1  - i \tan\delta_c) c_b/2$, whereas resonator density $\rho_0 = 10 \rho_b$ and compressibility $\beta_0 = 1/(\rho_0 c_0^2)$. 

The extinction $\sigma_\mathrm{ext}(\omega)$, absorption $\sigma_\mathrm{abs}(\omega)$ and scattering $\sigma_s(\omega)$ cross-sections obtained numerically using Eqs.~\eqref{eq:absorbed_power_volume_integral}--\eqref{eq:scattered_power_volume_integral} for the considered two-dimensional cylindrical resonator are shown in Fig.~\ref{Draft_Figure_2}. To compare with, we show in Fig.~\ref{Draft_Figure_2} cross-sections calculated within the Mie-theory framework with expansion coefficients (see details in Ref.~\cite{williams1999fourier} and Supplemental Material~\cite{Supplement_II}, Sec.~III). In order to perform the calculations, we consider the number of modes for decompositions~\eqref{eq:absorbed_power_volume_integral}--\eqref{eq:scattered_power_volume_integral} defined by the following parameters -- $kR < 35$ and maximal azimuthal mode number $m_{\max} = 30$. For this truncation, one may observe perfect correspondence between numerical simulations via the developed approach and the analytical Mie-theory solution, which leads to the difference in calculated spectra approaching zero.

\begin{figure} [t]
    \centering
    \includegraphics[width = \linewidth]{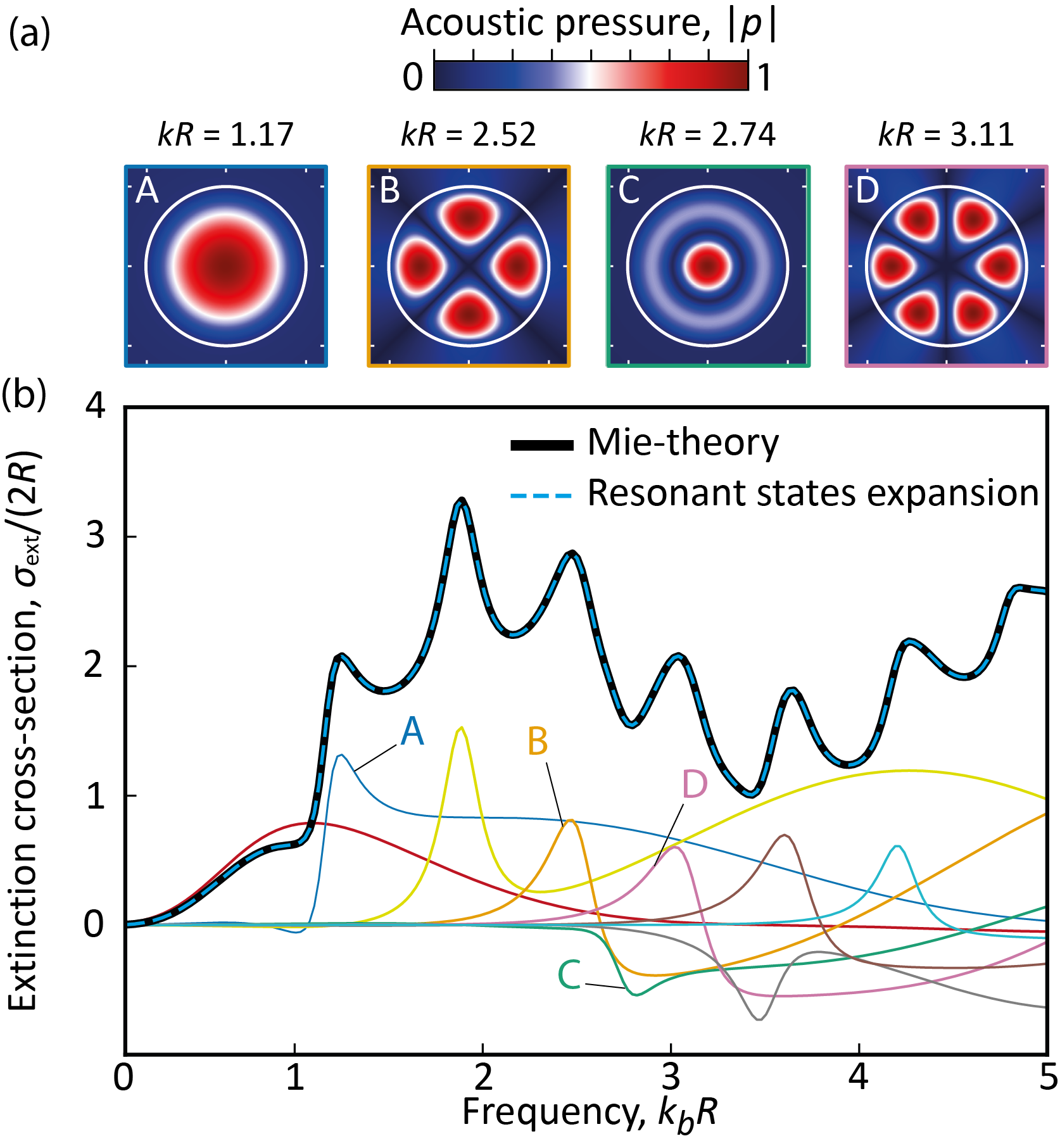}
    \caption{
    (a) Normalized pressure field modulus distributions for several RSs. The spectral positions of the RSs are listed above the field distributions and indicated in Fig.~\ref{Draft_Figure_4}(a).
    (b) The extinction $\sigma_\mathrm{ext}$ cross-section spectra of a 2D cylinder obtained analytically using the Mie theory (solid black line) and numerically using RSE and volume integration  [Eq.~\eqref{eq:extinction_power_volume_integral}] (dashed blue line). For more details, see Supplemental Material~\cite{Supplement_II}, Sec.~III. The coloured lines represent the contribution of individual RSs to the extinction cross-section spectrum. The geometric and material parameters are listed in the caption of Fig.~\ref{Draft_Figure_2}.}
    \label{Draft_Figure_3}
\end{figure}

Importantly, similar to the light scattering in optics~\cite{lalanne2018scattering,powell2017interference}, the extinction cross-section $\sigma_\mathrm{ext}(\omega)$ is a linear function with respect to the scattered field. But in fact, unlike optics, in RS scattered field decomposition, one observes linear decompositions for the pressure $p_s$ and sound velocity $\mathbf{v}_s$ fields with two material parameters $\rho$ and $\beta$ in acoustics. We demonstrate this extinction cross-section modal decomposition in Fig~\ref{Draft_Figure_3}. The excellent agreement between the RS extinction cross-section decomposition given by Eq.~\eqref{eq:extinction_power_volume_integral} and Mie-theory analytical results, obtained in Supplemental Material~\cite{Supplement_II}, Sec.~III, evidences that the developed formalism allows for a quantitative prediction of the extinction cross-section spectra with the precise contributions from every individual RS. Whereas individual resonator eigenstates can give a negative contribution to the extinction cross-section, the whole spectrum is positive at an arbitrary frequency.

\begin{figure} [t]
    \centering
    \includegraphics[width = \linewidth]{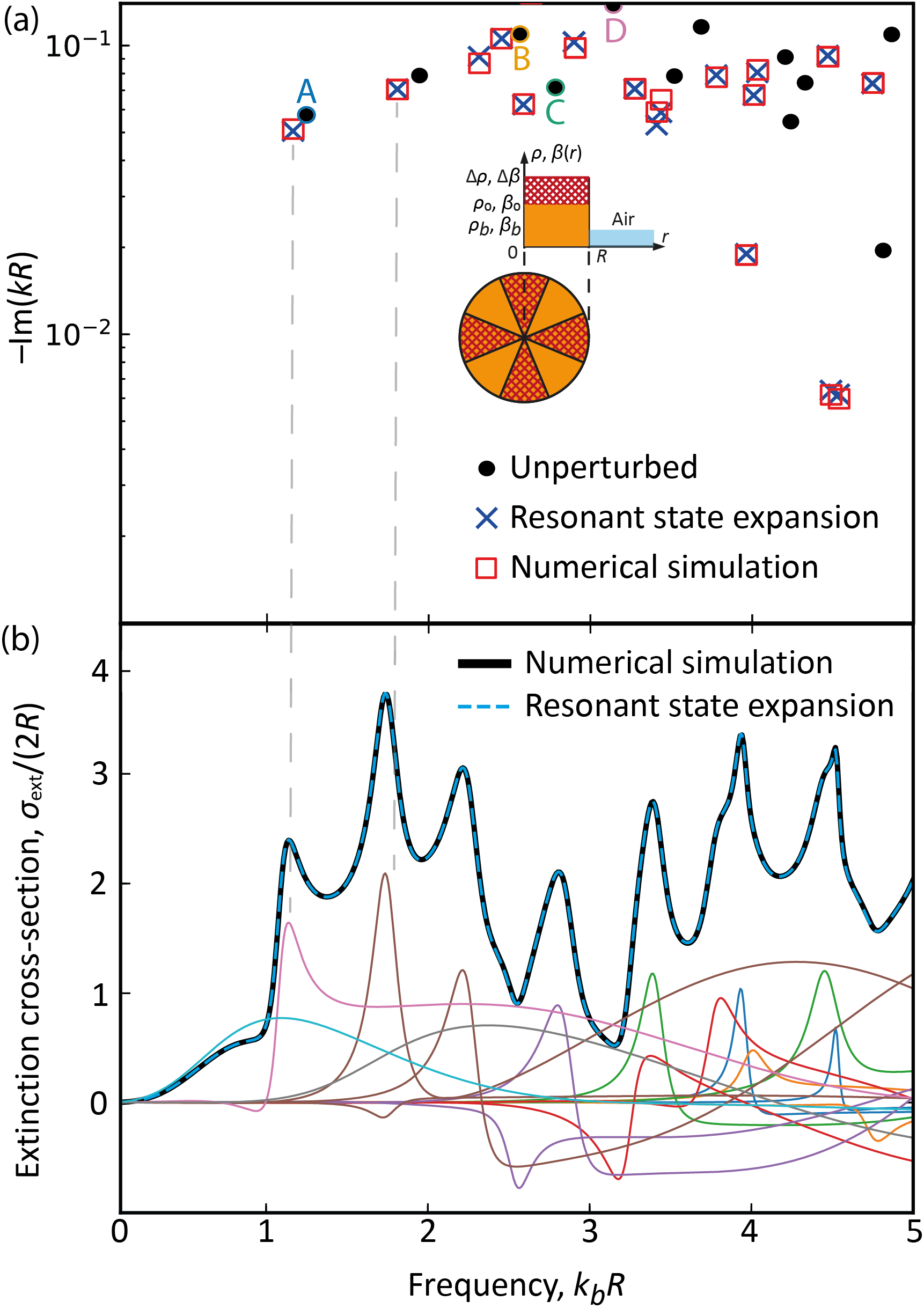}
    \caption{
    (a) Eigenfrequency spectra of the unperturbed (black circles) and $C_4$-sectoral perturbed cylindrical resonator system obtained by numerical simulation (open red squares) and RSE (blue crosses). The coloured circles with capital letters indicate the unperturbed RSs of the cylindrical resonator, which are listed in Fig.~\ref{Draft_Figure_3}(a).
    (b) The extinction $\sigma_\mathrm{ext}$ cross-section spectra of the sectoral perturbed system obtained within numerical simulations in COMSOL Multiphysics (solid black line) and obtained numerically for all RSs by Eq.~\eqref{eq:extinction_power_volume_integral} (dashed blue line). The coloured lines represent the contribution to the extinction cross-section spectra for each individual mode. Material parameters perturbations are $\Delta \rho = 0.1 \rho_0$ and $\Delta \beta = 0.2 \beta_0$.
    }
    \label{Draft_Figure_4}
\end{figure}

Next, we want to consider a sectorally perturbed cylindrical resonator with the symmetry $C_4$. The material parameters perturbation is given by $\Delta \rho(r,\varphi) = \Delta\rho S_4(\varphi) \Theta(R-r)$ and $\Delta \beta(r,\varphi) = \Delta\beta S_4(\varphi) \Theta(R-r)$, where $S_4(\varphi) = \sum_{q=0}^{3}\Pi_\theta \left(\varphi - \pi q/2 \right)$ is a dimensionless characteristic function describe material change angular distribution, $\Pi_\theta(\varphi) = 1$ for $|\varphi| \le \theta$ and zero otherwise. The eigenmodes of the perturbed system with perturbation aperture $2\theta = 45^\circ$ were calculated within the RSE framework with the material parameter perturbations $\Delta \rho = 0.1 \rho_0$ and $\Delta \beta = 0.2 \beta_0$, shown in Fig.~\ref{Draft_Figure_4}(a). The extinction $\sigma_\mathrm{ext}(\omega)$ cross-section spectra obtained numerically by Eq.~\eqref{eq:extinction_power_volume_integral} for the considered two-dimensional cylindrical resonator are shown in Fig.~\ref{Draft_Figure_4}(b). 

Unlike a homogeneous system, in this case, Mie theory does not work due to coupling between different cylindrical resonator eigenmodes. So, we perform a COMSOL Multiphysics numerical simulation to compare with the numerical calculation with Eq.~\eqref{eq:extinction_power_volume_integral} and show the results in Fig.~\ref{Draft_Figure_4}(b) for reference. As for the previous one, the contribution of the individual system eigenmode is also shown.

One may observe the excellent correspondence between the two numerical simulations, with the error approaching zero for the decomposition number of mode $N = 5694$. Moreover, we show the possibility of analyzing each mode scattering contribution individually. These statements make the developed approach an effective tool to analyze scattering on the acoustic resonator of an arbitrary shape. On the other hand, this approach can be implemented for more complicated systems and paired with the numerical calculations of eigenfrequencies and eigenmodes in any numerical software package, even COMSOL Multiphysics, to be specific.

\section{RSE Analysis of a Material-Programmed Acoustic Metaatom}
\label{sec:annular_metaatom}

The eigenvalue and scattering problems RSE formulations can be concatenated into a single modal design procedure for systems of an arbitrary shape. Consequently, the analytically solvable homogeneous cylinder, treated before, is not a unique reference system. It can be replaced by any open resonator for which a complete and correctly normalized RS basis is available, either analytically or numerically. The reference basis may therefore incorporate the principal geometrical features of a metaatom from the outset, while the RSE is used only to program its material parameter distribution.

\begin{figure} [t]
    \centering
    \includegraphics[width = \linewidth]{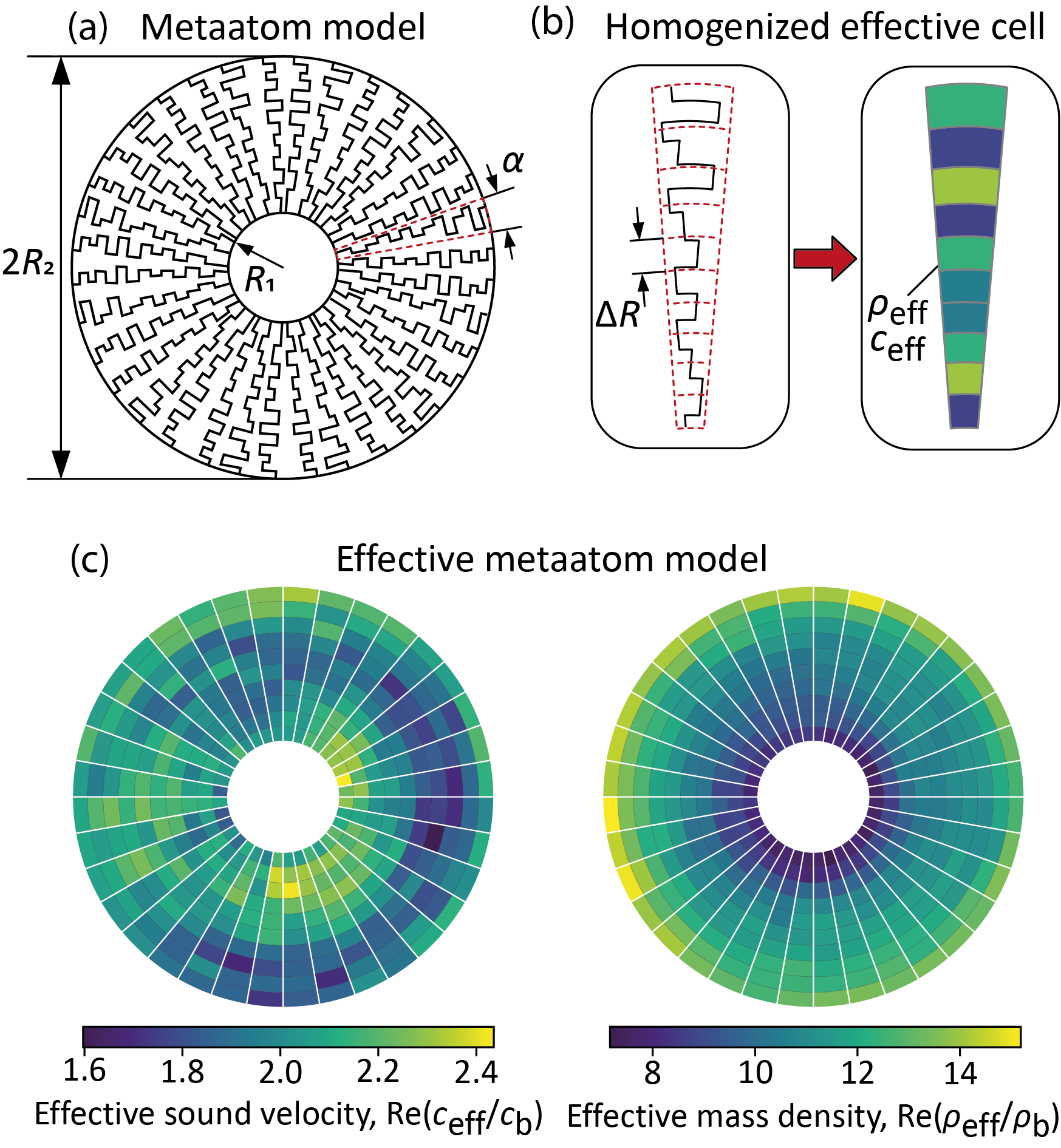}
    \caption{
    (a) Annular acoustic metaatom model with inner radius $R_1 = 35~\mathrm{mm}$ and outer radius $R_2 = 135~\mathrm{mm}$. Radial hard walls divide the annulus into $N_\varphi = 36$ air channels with aperture $\alpha = 10^\circ$. The sectors have no direct lateral flux inside the annulus but remain indirectly coupled through the common core and exterior air.
    (b) One acoustic channel divided into $N_r = 10$ radial sections with the same length $\Delta R = 10~\mathrm{mm}$ and its homogenized effective model with $N_r = 10$ sectoral unit cells. Each unit cell has its own effective sound velocity $c_\mathrm{eff}$ and mass density $\rho_\mathrm{eff}$.
    (c) Material distribution inside the effective annular metaatom model with $N_\varphi = 36$ and $N_r = 10$. Left: distribution of real sound velocity normalized to sound velocity of surrounding air $\mathrm{Re}(c_\mathrm{eff}/c_b)$. Right: distribution of real effective mass density normalized to mass density of surrounding air $\mathrm{Re}(\rho_\mathrm{eff}/\rho_b)$. White radial lines on both figures mark ideal hard-wall sectoral boundaries. See definition of all effective parameters in Supplemental Material~\cite{Supplement_II}, Sec.~IV~B
    }
    \label{Draft_Figure_5}
\end{figure}

In order to demonstrate this construction, as the last and the most complex case, we consider the space-coiled annular anisotropic metamaterial. Air with sound velocity $c_b = 343~\mathrm{m/s}$ and density $\rho_b = 1.2~\mathrm{kg/m}^3$ occupies the central region of inner radius $r < R_1$ and the exterior region with outer radius $r > R_2$ with the finite number $N_\varphi = 36$ of fan-shaped acoustic channels, each with the aperture $\alpha = 2\pi/N_\varphi = 10^\circ$ and stochastic structural parameters, shown in Fig.~\ref{Draft_Figure_5}(a). As an unperturbed system, we choose an annular resonator, for which the $C_M$-symmetry separates the unperturbed eigenvalue problem into $N_\varphi$ independent symmetry sectors with constant density $\rho_0 = 10 \rho_b$, sound velocity $c_0 = c_b/2$, and compressibility $\beta_0 = 1/\rho_0c_0^2$ in all sectors. Inside each fan-shaped channel, the pressure is expanded over Neumann angular functions, whereas the fields in the central and exterior air regions are expanded over regular Bessel and outgoing Hankel harmonics, respectively. Continuity of pressure $p$ and radial velocity $v_r$ at $R_1$ and $R_2$ gives a mode-matching matrix $\mathcal{F}_s(\omega)$ for each symmetry index $s = 0, 1, \ldots, N_\varphi - 1$. The complex resonant frequencies are therefore determined by
\begin{equation}
    \det \mathcal{F}_s(\omega_n) = 0,
    \label{eq:annular_secular_main}
\end{equation}
and the corresponding pressure and velocity fields are reconstructed from the null vectors of $\mathcal F_s(\omega_n)$ (see details in Refs.~\cite{xu2023argument,nagarsheth2021somenew} and Supplemental Material~\cite{Supplement_II}, Sec.~IV~A). The channels are separated by walls with finite width. Such a structure is of interest~\cite{jiang2019proposal,liang2012extreme,Zhao2018directional,li2025single-microphone,smagin2026unidirectional} due to the possibility of controlling wave sound velocity by variation of channel parameters, for example, its length and width. Notwithstanding the physical schematic containing finite-width partitions, the effective model used in the RSE and matched numerical simulations calculations employs ideal zero-thickness hard-wall boundaries. Each of the $N_r = 10$ radial cells is assigned the effective material values, as shown in Fig.~\ref{Draft_Figure_5}(b). As a result, the considered system is replaced by the effective model where each part of the acoustic channels is changed to an isotropic sectoral area with effective material parameters, as shown in Fig.~\ref{Draft_Figure_5}(c). Definitions, physical derivation, validity conditions, and numerical ranges of the effective parameters are given exclusively in Supplemental Material~\cite{Supplement_II}, Sec.~IV~B. The hard walls suppress lateral flux inside the annulus. The circular interfaces at $r = R_1$ and $r = R_2$ remain acoustically open, so that the RSs radiate both into the central cavity and into the external medium and possess complex eigenfrequencies. The hard-wall partitions are included exactly in the reference geometry rather than treated as a material perturbation.

The target metaatom is generated by subdividing every sector into $N_r = 10$ radial cells. In the cell $\Omega_{jq}$, the effective parameters are independently specified as $\rho_{jq}^{\mathrm{eff}}$, $\beta_{jq}^{\mathrm{eff}}$ and $c_{jq}^{\mathrm{eff}}$, derived in Supplemental Material~\cite{Supplement_II}, Sec.~IV~B. The material perturbation is
\begin{equation}
    \Delta \hat{\mathbb{P}}(r,\varphi) = \sum_{j=0}^{N_{\varphi}-1}\sum_{q=1}^{N_r}
    f_{jq}(r,\varphi)
    \begin{bmatrix}
        -\Delta\beta_{jq} & 0\\
        0 & \Delta\rho_{jq}
    \end{bmatrix},
    \label{eq:annular_material_perturbation}
\end{equation}
where $f_{jq}(r,\varphi) = \Theta_q(r) \Theta_j(\varphi)$ is the characteristic function of the unit cell that defines the effective material parameters' spatial distribution. It is expressed by radial $\Theta_q(r) = \Theta(r - r_{q-1}) - \Theta(r - r_{q})$ and sectoral $\Theta_j(\varphi) = \Theta(\varphi - \varphi_j) - \Theta(\varphi - \varphi_{j+1})$ Heaviside functions. Perturbations in Eq.~\eqref{eq:annular_material_perturbation} are given as $\Delta \beta_{jq} = \beta^{\mathrm{eff}}_{jq} - \beta_0$, and $\Delta \rho_{jq} = \rho^{\mathrm{eff}}_{jq} - \rho_0$. Thus, changing the $2N_{\varphi}N_r$ values of $\rho^{\mathrm{eff}}_{jq}$ and $c^{\mathrm{eff}}_{jq}$ requires only reassembly and diagonalization of the finite RSE matrix, meanwhile, the reference RSs and cell-resolved overlap tensors are calculated once.

The second step of the calculation uses the RSs of the perturbed system as the scattering basis of Eqs.~\eqref{eq:scattered_field} and~\eqref{eq:excitation_coefficients}. The physical scattered field relative to homogeneous air is the sum of the response of the fixed geometry and the material-induced field in Eq.~\eqref{eq:scattered_field}. The extinction, scattering, and absorption cross-sections then follow directly from Eqs.~\eqref{eq:cross_sections}, while the pressure $p$ and velocity $\mathbf{v}$ near fields are reconstructed from the same modal coefficients, derived in Supplemental Material~\cite{Supplement_II}, Sec.~IV~C.

\begin{figure} [t]
    \centering
    \includegraphics[width = \linewidth]{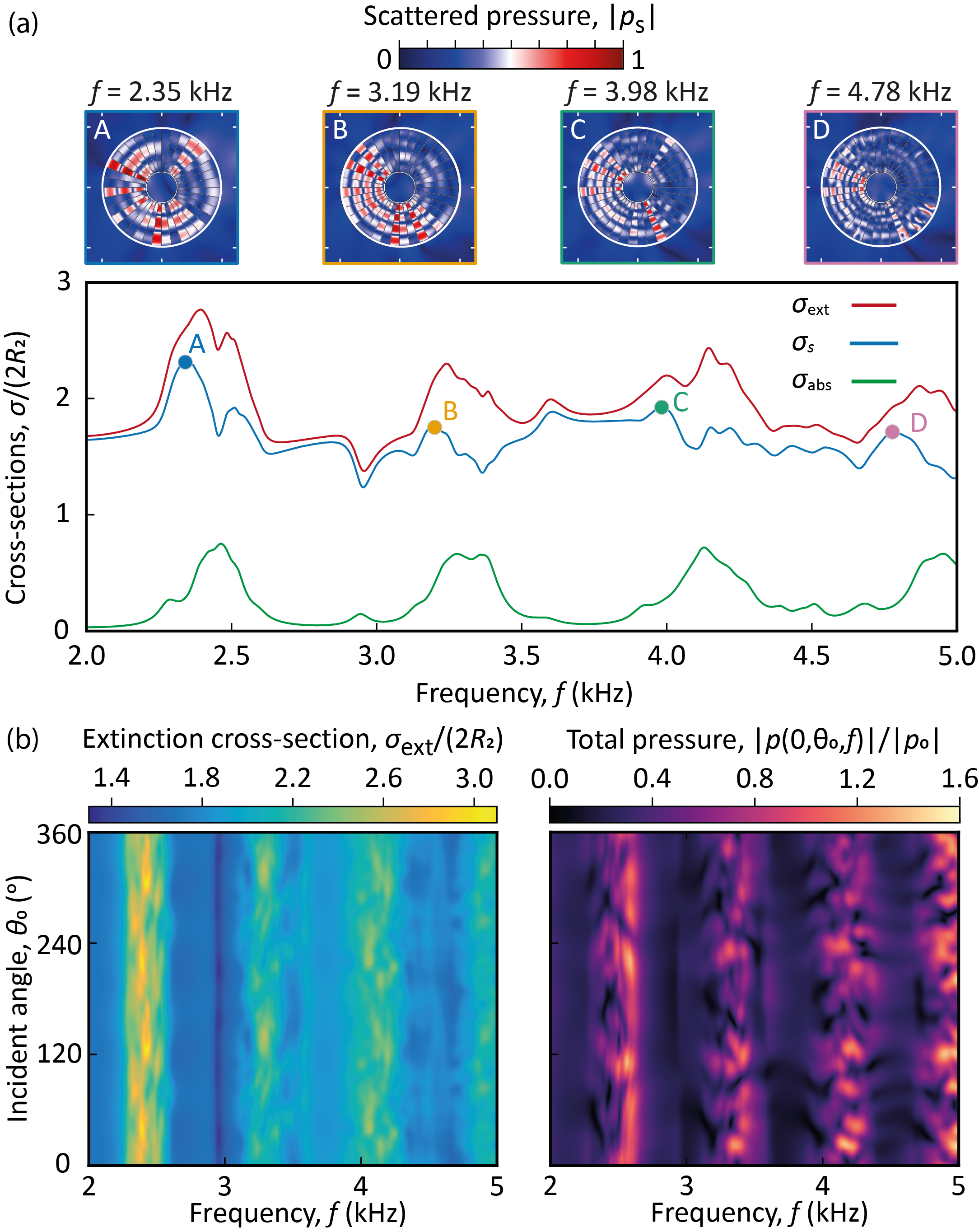}
    \caption{
    (a) Normalized extinction $\sigma_\mathrm{ext}$ (red line), scattering $\sigma_s$ (blue line) and absorption $\sigma_\mathrm{abs}$ (green line) cross-sections at incidence angle equals to $\theta_0 = 20^\circ$. The cross-sections are normalized to the outer diameter $2R_2$ of the considered annular metatom. The capitals A--D identify four refined local maxima of the scattering spectra $\sigma_s$ and show the corresponding full-domain normalized scattered-pressure modulus $|p_s|$ on one common physical scale in the panels A--D. The pressure field modulus distribution includes both the central cavity and radiating exterior.
    (b) Left: normalized extinction $\sigma_\mathrm{ext}/(2R_2)$ versus incidence angle $\theta_0$ and frequency $f$. Right: Normalized total pressure magnitude at the central point $|p(\mathbf{x} = 0, \theta_0, f)|/|p_0|$ versus incidence angle $\theta_0$ and frequency $f$.
    }
    \label{Draft_Figure_6}
\end{figure}

For a representative geometry inspired by the space-coiling acoustic camera of Ref.~\cite{jiang2019proposal}, we use $N_{\varphi} = 36$, $N_r = 10$, $R_1 = 35$~mm, and $R_2 = 135$~mm as discussed in Supplemental Material~\cite{Supplement_II}, Sec.~IV~D. Figure~\ref{Draft_Figure_6}(a) shows normalized calculated within the RSE framework extinction $\sigma_\mathrm{ext}$, scattering $\sigma_s$ and absorption $\sigma_\mathrm{abs}$ cross-sections. The scattering spectra contain several local scattering maxima denoted by capitals A--D throughout 2–-5 kHz, for which the distributions of the scattered pressure modulus $|p_s|$ are also shown in Fig.~\ref{Draft_Figure_6}(a). The A--D near fields occupy different channel subsets and possess different radiation patterns, demonstrating frequency-selective routing by the cell-wise material. The angle-resolved extinction and the central-pressure response in Fig.~\ref{Draft_Figure_6}(b) are complementary observables. Indeed, a large extinction $\sigma_\mathrm{ext}$ at an arbitrary frequency does not in general imply a large local pressure modulus $|p|$ at the center of the whole system where $\mathbf{x} = 0$. For the truncation, the amplitude of the incident pressure wave is $p_0 = 1~\mathrm{Pa}$. Finally, we compare the simulation results of the model developed on the RSE basis with the COMSOL Multiphysics model, which is used for independent validation of the cross-section spectrum and near-field pressure distribution, and examine the comparison in detail in the Supplemental Material~\cite{Supplement_II}, Sec.~IV~D.

This annular metaatom example closes the modal design loop. First, a reference RS basis is chosen to contain the dominant topology and boundary conditions of the intended metaatom. Second, treating the eigenvalue problem with the RSE maps this basis to a large family of material-programmed structures. Third, treating the scattering problem with the RSE maps the resulting complex eigenmodes to measurable spectra and fields. The only essential requirements on the reference basis are the outgoing-wave condition, generalized normalization, and completeness after all relevant discrete and continuum contributions are included. The same procedure can therefore be applied to annular, labyrinthine, Helmholtz-type, numerically generated, and other open acoustic resonators, substantially extending the RSE beyond the homogeneous cylindrical reference system.

\section{Conclusion}

In this work, we developed the RSE framework for acoustic wave scattering by open resonators. The proposed approach represents the scattered acoustic field as an expansion over the RSs of the system and expresses the excitation coefficients through overlap integrals between the incident field, the material perturbation, and the eigenmodes of the resonator. This formulation provides a direct link between the spectral response of an acoustic structure and its individual resonant modes. The developed formalism was applied to a two-dimensional cylindrical acoustic resonator. For the homogeneous axisymmetric case, the extinction, scattering, and absorption cross-sections calculated using the RSE were shown to be in excellent agreement with the analytical Mie-theory solution. This agreement validates the proposed method and confirms that the RS basis correctly reconstructs the scattering response of the open acoustic system. A key advantage of the method is its ability to decompose the total extinction spectrum into contributions from individual RSs. This makes it possible not only to reproduce the full scattering spectrum, but also to identify which eigenmodes are responsible for particular spectral peaks and resonant features. Such a modal interpretation is especially important for open non-Hermitian acoustic systems, where resonances have complex eigenfrequencies and radiative losses.

The approach was further tested for a cylindrical resonator with a sectoral material perturbation, where the rotational symmetry is reduced, and different azimuthal modes become coupled. In this case, the standard Mie-theory description is no longer applicable. The RSE, however, successfully described the perturbed system and demonstrated excellent agreement with direct numerical simulations performed in COMSOL Multiphysics. This confirms that the method can be used for acoustic resonators with nontrivial material distributions and reduced symmetry. As a final application, we utilize the RSE for a material-programmed sectoral hard-wall annular metaatom. The driven target problem is solved directly in a converged angular finite section, and its fixed-angle scattering spectrum and near fields are validated against a matched COMSOL Multiphysics model. Thus, the developed RSE provides an efficient semi-analytical tool for studying acoustic scattering in open resonators of complex geometry. It combines the accuracy of numerical simulations with the physical transparency of modal analysis, allowing one to calculate scattering spectra, absorption, extinction, and individual modal contributions. The proposed framework can be further extended to more complicated acoustic metamaterials, labyrinthine resonators, lossy and dispersive media, and non-Hermitian acoustic systems with engineered modal interactions.

\begin{acknowledgments}
The authors acknowledge financial support from the Russian Science Foundation (25-79-31027).
\end{acknowledgments}

\bibliography{references_PRB_complete}

\end{document}